\documentclass[twocolumn,showpacs,showkeys]{revtex4}

\newcommand{\bastar}{\begin{eqnarray*}}
\newcommand{\eastar}{\end{eqnarray*}}
\newskip\humongous \humongous=0pt plus 1000pt minus 1000pt

\newif\ifdtup

\relax
\newcommand{\bea}{\begin{eqnarray}}
\newcommand{\eea}{\end{eqnarray}}
\newcommand{\dfrac}{\displaystyle\frac}
\newcommand{\nn}{\nonumber}
\newcommand{\tF}{{\tilde F}}
\newcommand{\bmu}{{\bar \mu}}
\newcommand{\bare}{{\bar e}}
\begin{document}
\noindent {\bf Cho and Pak Reply}: A recent Comment \cite{lamm} 
has criticized the logarithmic correction term of 
the effective action of QED in our Letter \cite{cho1},
\bea
&{\cal L}_{eff} =
-\dfrac{a^2-b^2}{2e^2}{\Big (}1-\dfrac{e^2}{12\pi^2}
\ln\dfrac{m^2}{\mu^2}{\Big)} \nn\\
&- \dfrac{ab}{4\pi^3} \sum_{n=1}^{\infty}\dfrac{1}{n}
{\Big [}\coth(\dfrac{n \pi b}{a}){\Big (} {\rm ci}(\dfrac{n \pi m^2}{a})
\cos(\dfrac{n \pi m^2}{a}) \nn\\
&+{\rm si}(\dfrac{n \pi m^2}{a}) \sin(\dfrac{n \pi m^2}{a}){\Big )} \nn\\
&-\dfrac{1}{2} \coth (\dfrac{n \pi a}{b})
{\Big (} \exp(\dfrac{n \pi m^2}{b})
{\rm Ei}(-\dfrac{n \pi m^2}{b}) \nn\\
&+ \exp(-\dfrac{n \pi m^2}{b}){\rm Ei}(\dfrac{n \pi m^2}{b}
-i\epsilon){\Big )}{\Big ]},
\eea
where $\mu$ is the subtraction parameter and
\bea
&a = \dfrac{e}{2} \sqrt {\sqrt {F^4 + (F \tF)^2} + F^2}, \nn\\
&b = \dfrac{e}{2} \sqrt {\sqrt {F^4 + (F \tF)^2} - F^2}. \nn
\eea
The Comment claims that ``the logarithmic correction term found by Cho and 
Pak vanishes when the final result is written in terms of the finite,
renormalized, physical electron charge'', asserting that ``these terms
do not appear if on-mass shell renormalization is used''.  

\indent We have no intention to dispute this claim, 
because this is a simple reiteration of what everybody knows.
What we like to point out here is that this criticism is based on
the confusion of the regularization with
the renormalization. Our Letter \cite{cho1} 
deals only with the regularized
effective action, {\it not} the renormalized one. And
the logarithmic term in Eq. (1) contains an important
piece of information, the subtraction dependence of 
the regularized effective action. The renormalization
(and the renormalization group invariance) of the effective action
has already been fully discussed in a separate paper \cite{cho2}, 
which was quoted in Ref. [12] of \cite{cho1}. 
Here we simply note that the logarithmic term
{\it does} disappear if we use the mass-shell subtraction $\mu=m$.
This is exactly what the Comment asserts. 
This nullifies the critisism of the Comment even without
the renormalization of the effective action.

We also stand by our remark in Ref. [9]
of \cite{cho1}: An honest regularization of the divergent integral
expression of the QED effective action must produce 
both the imaginary part and the logarithmic term.
Only after the renormalization does the 
logarithmic term disappear, as we discussed in \cite{cho2}.
Furthermore, we remark that the logarithmic term 
plays an important role in proving the fact that the QED effective
action has no infra-red divergence in the massless limit when $ab=0$.
This is evident from Eqs. (10), (12), and (14) of \cite{cho1}.

One might wonder why we did not use the simple 
``mass shell renormalization'' in our Letter [2]. 
The reason is that one can not use the mass shell renormalization 
in a massless theory, because it can not control the infra-red
divergence properly. And if one wants to discuss the massless limit
of QED, one must do the the subtraction dependent regularization first.
This point becomes more important when one tries to calculate
the effective action of QCD \cite{cho3}. 

The renormalization 
of the effective action gives us an 
unexpected surprise. To renormalize the effective action, one need to
define the running coupling $\bare (\bmu)$ by \cite{cho2} 
\bea
\dfrac{d^2 V_{eff}}{d a^2}{\Big |}_{a = {\bar \mu}^2} 
= \dfrac{1}{\bare^2},
\eea
where $V_{eff}$ is the effective potential obtained from Eq. (1). 
A remarkable point here is that this definition produces
the running coupling which is different from what one obtains 
from the perturbative calculation \cite{cho2}. This is surprising, because 
in QCD the above definition and the perturbative calculation 
produce an identical result \cite{cho3}. Only in QED do we have
this discrepancy. A possible interpretation of 
the origin of this difference is discussed
in \cite{cho2}.

Note Added: There is a typological mistake in \cite{cho1}.
The RHS of Eq. (24) in \cite{cho1} should have an overall minus
sign.
\vspace{0.5 cm}

\noindent Y.M. Cho$^1$ and D.G. Pak$^2$ \\
\indent$^1$School of Physics \\ 
\indent ~Seoul National University \\
\indent ~Seoul 151-742, Korea \\
\indent$^2$Institute of Applied Physics \\
\indent ~Uzbekistan National University \\
\indent ~Tashkent 700-174, Uzbekistan \\
\vspace{0.5 cm}

\noindent DOI: 

\end{document}